\documentclass[prc,twocolumn,epsfig]{revtex4}
\usepackage{graphics}
\usepackage{epsfig}
\usepackage{amsfonts}
\usepackage{amsmath}
\usepackage{bm}

\usepackage{color}
\begin{document}
\title{Individual low-energy E1 toroidal and compression states in light nuclei:
deformation effect, spectroscopy and interpretation}
\author{
 V.O. Nesterenko$^{1,2,3}$, J. Kvasil$^4$, A. Repko $^{5}$, and P.-G. Reinhard$^6$}
\affiliation{$^1$ Laboratory of Theoretical Physics,
  Joint Institute for Nuclear Research, Dubna, Moscow region, 141980, Russia}
 \affiliation{$^2$
 State University "Dubna", Dubna, Moscow Region, 141980, Russia}
 \affiliation{$^3$
 Moscow Institute of Physics and Technology, Dolgoprudny, Moscow region, 141701, Russia}
 \affiliation{$^4$
 Institute of Particle and Nuclear Physics, Charles University, CZ-18000, Praha 8, Czech Republic}
  \affiliation{$^5$
  Institute of Physics, Slovak Academy of Sciences, 84511, Bratislava, Slovakia}
  \affiliation{$^6$
 Institut f\"ur Theoretische Physik II, Universit\"at Erlangen, D-91058, Erlangen, Germany}

\date{\today}

\begin{abstract}
 The existence of individual low-energy E1 toroidal and
  compression states (TS and CS) in $^{24}$Mg was predicted recently
  in the framework of quasiparticle random-phase-approximation (QRPA)
  model with Skyrme forces. It was shown that the strong axial
  deformation of $^{24}$Mg is crucial to downshift the toroidal
  strength to the low-energy region and thus make the TS the lowest
  E1(K=1) dipole state. In this study, we explain this result by
  simple mean-field arguments.  Comparing TS in two strongly axial
  nuclei, $^{24}$Mg and $^{20}$Ne, we show that the lowest TS
  is not not a universal phenomenon but rather a peculiarity
  of $^{24}$Mg. The spectroscopy of TS and CS is analyzed and
  some additional interpretation of these states is suggested.
\end{abstract}
\pacs{21.10.-k,21.60.-n,21.60.Jz}

\maketitle

\section{Introduction}
\label{intro}

In a recent publication \cite{Ne_PRL18}, we have predicted the
occurence of individual low-energy E1 toroidal and compressional states
(TS and CS) in $^{24}$Mg. The calculations were performed within the
quasiparticle random-phase-approximation (QRPA) method using several
Skyrme forces. This prediction opens a new promising path in
exploration of toroidal excitations.  Previously the nuclear toroidal
mode was mainly studied in terms of the giant isoscalar (T=0) toroidal
dipole resonance (TDR), see e.g.
\cite{Se81,Bas93,Bal94,Mis06,Ry02,Co00,Vr02,Pa07,Kv11,Rep13,Rei14,NePAN16,Rep17}
and references therein. However the experimental observation and
identification of the TDR is plagued with serious troubles. The
resonance is usually masked by other multipole modes (including dipole
excitations of non-toroidal nature) located at the same energy
region. As a result, even the most reliable $(\alpha,\alpha')$
experimental data for E1(T=0) TDR \cite{YoungexpZr,Uch04} can be
disputed, see the detailed discussion in Ref. \cite{Rep17}.  In this
respect, the individual low-energy E1(T=0) TS in light nuclei have
significant advantages. As shown \cite{Ne_PRL18}, the TS in axially
deformed $^{24}$Mg should appear as the lowest (E=7.92 MeV) dipole
state with K=1 (K is the projection of the total angular momentum to
the symmetry z-axis). It is well separated from the neighbouring
dipole states, which simplifies its experimental identification and
exploration as compared to the TDR. Low-energy TS were also predicted
in other light deformed nuclei, see e.g. the AMD (antisymmetrized
molecular dynamics) calculations for $^{10}$Be \cite{Kanada17}.

As demonstrated in \cite{Ne_PRL18}, the TS in $^{24}$Mg becomes the
lowest E1(K=1) state due to the large deformation-induced downshift of
its excitation energy. So just the large axial quadrupole deformation
makes the TS an individual mode, well separated from other states.
This finding rises two questions. 1) How to explain this
deformation effect by simple mean-field arguments? 2) How universal is
this effect in strongly deformed light nuclei? These questions are
addressed in the present study where we compare the impact
of deformation in two strongly deformed N=Z nuclei, $^{24}$Mg and
$^{20}$Ne. For completeness, both vortical TS and irrotational CS
are considered. It will be shown that the significant deformation-induced
energy downshift is pertinent to both of $^{24}$Mg and $^{20}$Ne.  However,
in  $^{20}$Ne, the TS is not the lowest dipole K=1 state anymore.
Moreover, in $^{20}$Ne, TS lies higher than CS. This means
that $^{24}$Mg is perhaps rather unique light nucleus where the
conditions for discrimination of the TS are most convenient.

Besides, we consider spectroscopic properties of TS and CS and suggest their
interpretation in terms of low-energy T=0 dipole states with isospin-forbidden
E1 ransition (for TS) and octupole mode (for CS).

\section{Calculation scheme}
\label{sec-2}

The calculations are performed within self-consistent QRPA based on
the Skyrme functional \cite{Ben03}. We use the Skyrme parametrization
SLy6 \cite{SLy6} as in the previous study \cite{Ne_PRL18}.  The QRPA
code for axial nuclei \cite{Repcode} exploits a 2D coordinate-space
mesh in cylindrical coordinates.  The single-particle spectrum is
taken from the bottom of the potential well up to +55 MeV in the
contiuum. The equilibrium deformation is calculated by minimization of
the total energy. Calculations with non-equilibrium deformations are
performed using quadrupole constrained Hartree-Fock.  The volume
monopole pairing is treated at the BCS level \cite{Rep17}. The QRPA
uses two-quasiparticle (2qp) basis with excitation energies up
to $\sim$ 100 MeV. The basis includes $\approx$ 1700-1900 (K=0) and
$\approx$ 3200-3600 (K=1) states. The Thomas-Reiche-Kuhn sum rule
\cite{Ring_book,Ne08} and isoscalar dipole energy-weighted sum rule
\cite{Harakeh_book_01} are exhausted by 100$\%$ and 97$\%$,
respectively.

Since vortical toroidal and irrotational compressional modes are
coupled \cite{Vr02,Pa07,Kv11}, we inspect both TS and CS. The
toroidal and compression dipole responses (reduced transition probabilities) are
\begin{equation}\label{BE1K}
B_{\nu}(E1K, \alpha)=(2-\delta_{K,0})|\:\langle\nu|\:\hat{M}_{\alpha}(\;E1K)\:|0\rangle \:|^2
\end{equation}
where $|\nu\rangle$ is the  wave function of the $\nu$-th QRPA dipole state.
The  toroidal ($\alpha$ = tor) and compressional ($\alpha$=com) transition
operators  are \cite{Kv11,Rep13,Ne_PRL18}
\begin{equation}\label{TM_curl}
\hat{M}_{\text{tor}}(E1K) = \frac{-1}{10 \sqrt{2}c} \int d^3r r [r^2+d^s+d^a_K]
{\bf Y}_{11K} \cdot ( \bf{\nabla} \! \times \! \hat{\bf j}),
\end{equation}
\begin{equation}\label{CM_div}
\hat{M}_{\text{com}}(E1K) =  \frac{-i}{10c}\int d^3r r[r^2+d^s-2d^a_K]
Y_{1K} (\bf{\nabla} \cdot \hat{\bf j}),
\end{equation}
where $\hat{\bf j}(\bf r)$ is operator of the nuclear current;
${\bf Y}_{11K}(\hat{\bf r})$ and $Y_{1K}(\hat{\bf r})$ are vector
and ordinary spherical harmonics;
$d^s= - 5/3 \langle r^2\rangle_0$ is the center-of-mass correction (c.m.c.)
in spherical nuclei \cite{Kv11};
$d^a_K = \sqrt{4\pi/45}\langle r^2 Y_{20} \rangle_0 (3\delta_{K,0}-1)$
is the additional c.m.c. in axial deformed nuclei within the prescription
\cite{Kvsp}. The average values mean $\langle f \rangle_0 = \int\:d^3r f \:\rho_0 /A$
where  $\rho_0$ is the g.s. density. We have checked that these c.m.c. accurately
remove spurious admixtures.

The toroidal operator (\ref{TM_curl})  with the curl
$\bf{\nabla}\!\times\!\hat{\bf j}$ is vortical while the compression operator
(\ref{CM_div}) with the divergence
$\bf{\nabla}\!\cdot\!\hat{\bf j}$ is irrotational.
Using the continuity equation, the current-dependent operator (\ref{CM_div})
can be transformed \cite{Kv11} to the familiar density-dependent form \cite{Harakeh_book_01}
$
\hat M'_{\text{com}}(E1K) = 1/10 \int d^3r r \hat{\rho} [r^2+d^s-2d^a_K ] Y_{1K}
$
with $\hat{\rho}(\bf r)$ being the density operator.

The isoscalar (T=0) nuclear current $\hat{\bf j}$ includes the convection
part $\bold {j}^q_c$ (with effective charges  $e^{\rm n,p}_{\rm eff}=0.5$)
and magnetization (spin) part $\bold {j}^q_m$ (with g-factors
$g^{\rm n,p}_s=(\bar{g}_{s}^n + \bar{g}_{s}^p) \eta /2 = 0.88 \eta$
where $\bar{g}^{\rm n,p}_s$ are bare g-factors and $\eta$ =0.7 is the
quenching) \cite{Kv11}. T=0 responses are relevant when considering
TS and CS in isoscalar reactions like $(\alpha,\alpha')$.
The  fields of the convection nuclear current are calculated as the current transition
densities (CTD) $\delta \bold {j}^q_c = \langle \nu| \hat{\bold j}^q_c|0\rangle$.

\section{Results and discussion}
\subsection{$^{24}$Mg}

First of all we recall the results \cite{Ne_PRL18} concerning the
deformation-induced energy downshift for the toroidal and
compressional responses in $^{24}$Mg. In figure 1, the
responses (\ref{BE1K}) for K=1 dipole states are depicted for
different deformations \cite{Ne_PRL18}, including the calculated
equilibrium deformation $\beta$=0.536 (which is rather close to the
experimental value $\beta_{\rm{exp}}$=0.605).  Figure 1
shows that increase of the deformation downshifts the toroidal
strength yielding eventually a remarked toroidal state with energy
7.92 MeV, the lowest in the dipole spectrum. The compressional K=1
strength which is much weaker than the toroidal one is also
downshifted with increasing deformation. The plots e) and f) show
that the 7.92 MeV state, being mainly toroidal, has also a minor
irrotational compression admixture.

\begin{figure} 
\includegraphics[width=8cm]{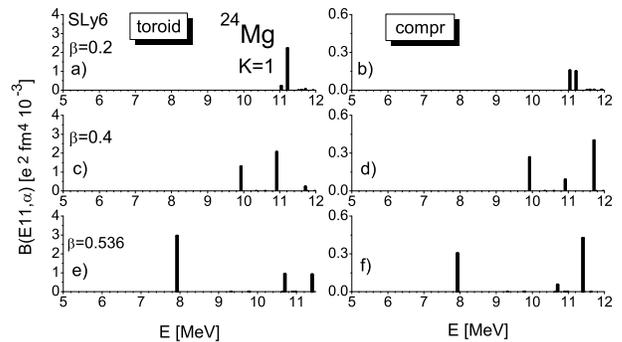}
\caption{
T=0 toroidal (left) and compression (right)
B(E11)-responses for K=1 states in $^{24}$Mg, calculated
for deformations $\beta$=0.2 (upper), 0.4 (middle),
and equilibrium deformation 0.536 (bottom) \protect\cite{Ne_PRL18}.
Note different scales in the right and left plots.}
\label{fig-1}       
\end{figure}

\begin{figure}
\includegraphics[width=8cm,clip]{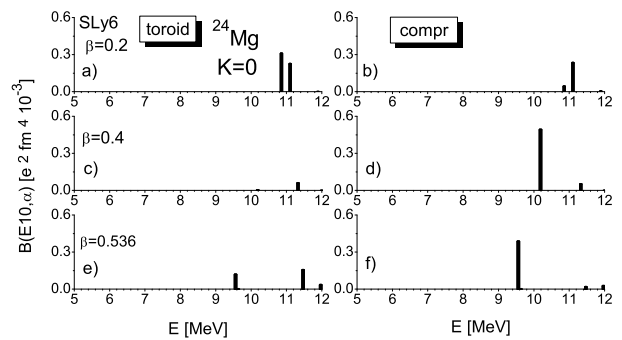}
\caption{The same in Fig. 1 but for K=0 dipole states \cite{Ne_PRL18}.
}
\label{fig-2}       
\end{figure}

Figure 2 exhibits similar responses for K=0 states
\cite{Ne_PRL18}. Here we see a definite downshift mainly
for the compressional strength. At  $\beta$=0.536, it
yields a noticeable peak at 9.56 MeV,
constituting the lowest K=0 state. Being mainly compressional,
this state has also a small toroidal fraction.
In K=0 channel, the toroidal strength is  much weaker than  in K=1 case.

Altogether, figures 1 and 2 show that, in
accordance with the previous studies for low-energy dipole spectra in
rare-earth deformed nuclei \cite{Rep17,KvYb},  the toroidal (compressional) mode
 dominates in K=1 (K=0) strength. In $^{24}$Mg,
both modes exhibit a significant deformation-induced downshift.
As a result, they acquire the lowest energies in their K-channels and
become well separated from higher dipole states. This can essentially
facilitate their experimental discrimination.

To understand the impact of deformation, it is instructive to consider
the structure of the toroidal K=1 7.92-MeV state and compressional K=0
9.56-MeV state.  Table I shows that the toroidal 7.92-MeV state is
mainly formed by two (proton and neutron) 2qp components of the same
content. Altogether they exhaust 93$\%$ of the states norm. In the
more collective 9.56 MeV state, two main 2qp components exhaust
70$\%$. In both cases, the major components are quite large and so
should dominate features of the states.

In figure 3, the deformation dependence of
the energies of single-particle states from the major 2qp components
is depicted. Only proton states are considered  since in N=Z nuclei
the neutron states exhibit a similar behavior. We see that, for $\beta \ge $ 0.4,
the deformation growth yields a rapid decrease  of [330]$\uparrow$ -energy,
significant increase of [101]$\downarrow$ -energy  and a relatively small change
of the energies for [211]$\uparrow$ and [211]$\downarrow$ .
This leads to decrease of the energies
of 2qp configurations pp[211]$\uparrow$-[330]$\uparrow$
and pp[211]$\downarrow$-[101]$\downarrow$ with $\beta$, which in turn explains
the downshift of the toroidal and compressional modes at large deformations.
\begin{table}
\caption{Properties of the lowest QRPA dipole states in $^{24}$Mg at
$\beta$=0.536.  The columns include: the excitation energy E (in MeV),
K-value, transition rate $B(E3K,0^+_{\rm{gs}}0 \to 3^-K)$ (in W.u.),
two main 2qp components (in Nilsson asymptotic quantum numbers),
position (F-pos) of the single-particle levels relative to the Fermi (F)
level, and contributions of 2qp components $\mathcal{N}$ to the state norm.}
\label{tab-1}       
\begin{tabular}{llllll}
\hline
E    & K & B(E3K)& main 2qp comp & F-pos & $\mathcal{N}$ \\\hline
7.92 & 1 & 11.7 & pp[211]$\uparrow$-[330]$\uparrow$ & F, F+4 & 0.54 \\
     &   &      &nn[211]$\uparrow$-[330]$\uparrow$ & F, F+4 & 0.39\\
9.56 & 0 &  17.0 &pp[211]$\downarrow$-[101]$\downarrow$ & F-2, F+2 & 0.39  \\
     &   &      &nn[211]$\downarrow$-[101]$\downarrow$ & F-2, F+2 & 0.31\\
\hline
\end{tabular}
\end{table}

\begin{figure} 
\includegraphics[width=7cm,clip]{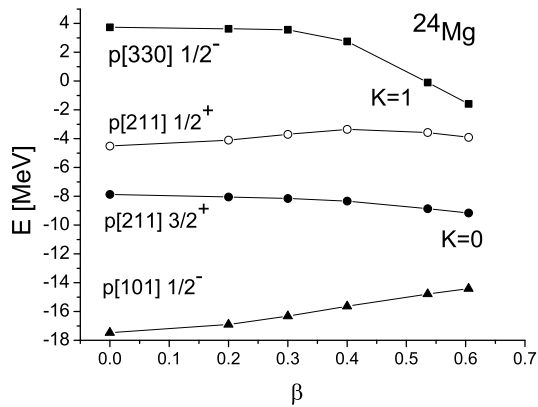}
\caption{Energies of proton single-particle states in
$^{24}$Mg.}
\label{fig-3}       
\end{figure}

Since the energy downshift in $^{24}$Mg is mainly determined by the
deformation dependence of particular single-particle states in
near Fermi energy, this effect is not universal.
In other words, TS and CS in other deformed light nuclei
are not necessarily the lowest dipole states.
To demonstrate this, we compare below the results for $^{24}$Mg  with those
for $^{20}$Ne, another N=Z nucleus with the strong quadrupole deformation.

\subsection{$^{20}$Ne}

For $^{20}$Ne, our SLy6 calculations give the equilibrium deformation
$\beta$= 0.56 which is considerably smaller than the experimental
value $\beta_{\rm{exp}}$=0.72 \cite{bnl}.  Note that
$\beta_{\rm{exp}}$ is determined by B(E20) transition $I^{\pi}K=0^+0
\to 2^+0$ in the ground state band and so
acquires large contributions from
dynamical correlations, especially in soft nuclei. As a result, the
obtained $\beta_{\rm{exp}}$ are allowed to be larger than the actual
"static" equilibrium deformation. Below we present the responses for
both axial deformations, $\beta$= 0.56 and 0.72.

\begin{figure}[h] 
\includegraphics[width=8cm,clip]{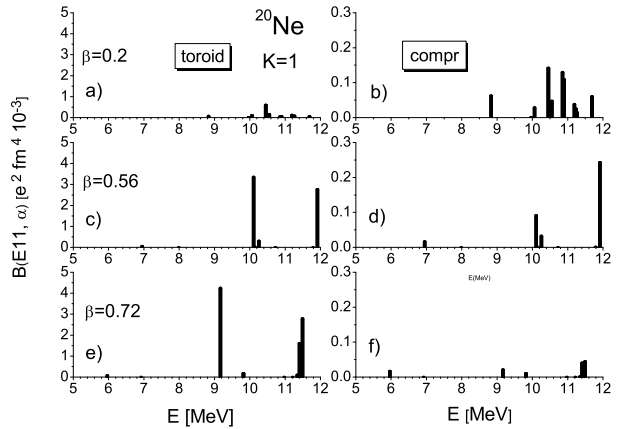}
\caption{T=0 toroidal (left) and compressional (right)
B(E11)-responses for K=1 states in $^{20}$Ne, calculated
for deformations $\beta$=0.2 (upper), 0.56 (middle),
and 0.72 (bottom). Note different scales in the left and right plots.}
\label{fig-4}       
\end{figure}
\begin{figure}[h] 
\includegraphics[width=8cm,clip]{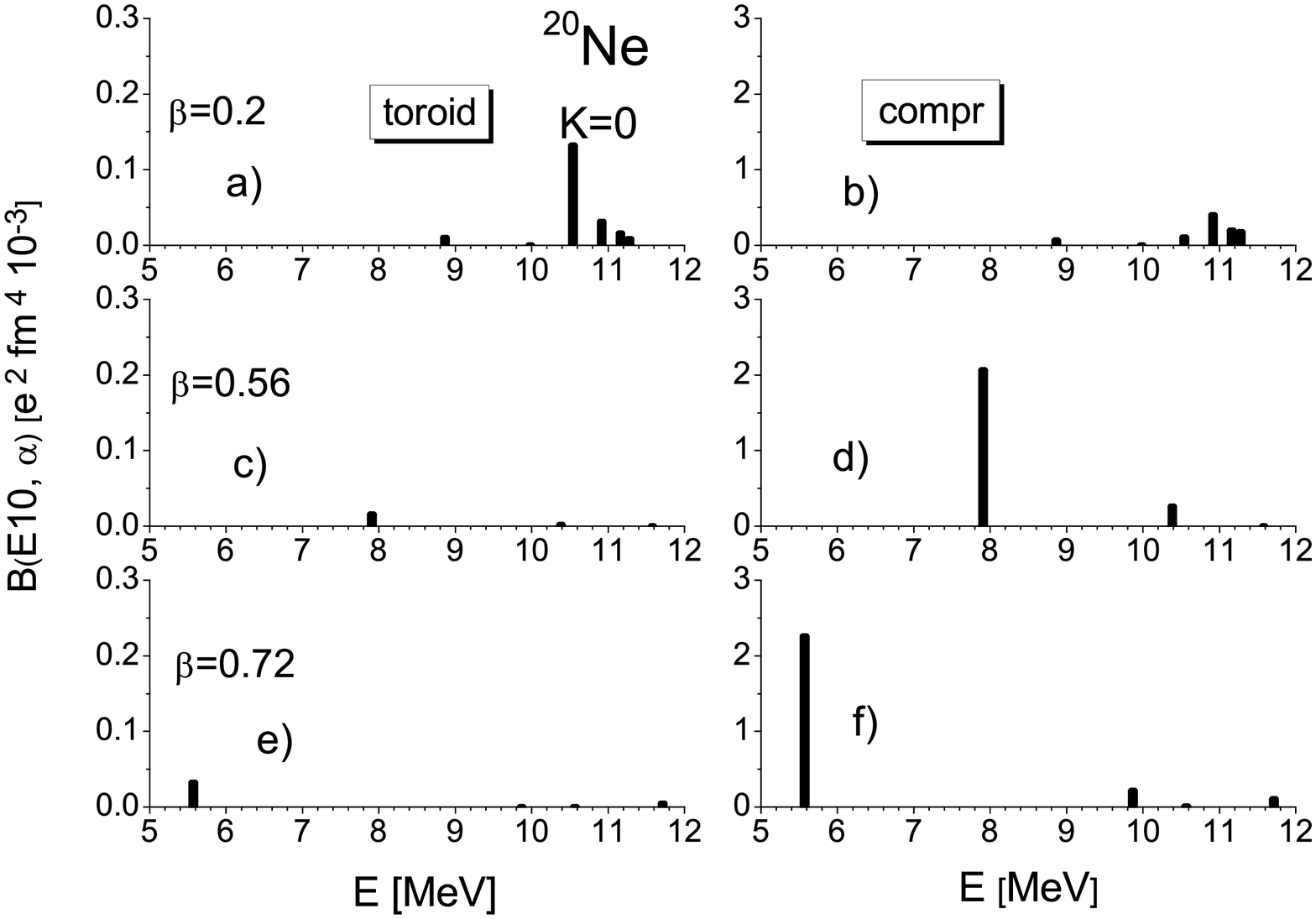}
\caption{The same as in Fig. 4 but for K=0 states.
Note different scales in the left and right plots.}
\label{fig-5}       
\end{figure}

Figures 4 and 5 show the toroidal and
compressional responses in $^{20}$Ne for K=1 and K=0 sates
at deformations $\beta$=0.2, 0.56 and 0.72. We see that,
in accordance with the results for $^{24}$Mg, the toroidal mode
dominates in K=1 strength while the compressional mode is major for
K=0. The responses are peaked in certain states. At
$\beta$=0.56, there are the strong toroidal K=1 peak at 10.11 MeV and
compressional K=0 peak at 7.91 MeV. The toroidal and compressional
nature of these states is confirmed by the pattern of the convective
current exhibited in figures 6 and 7.
\begin{figure}
\includegraphics[width=7cm]{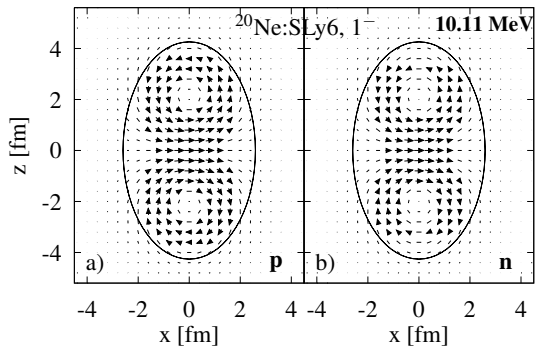}
\caption{Proton (left) and neutron (right) QRPA (SLy6) convective
currents (CTD) in z-x (y=0) plane for the K=1 10.1-MeV state in $^{20}$Ne.
Magnitude of the currents is determined by arrow
lengths in arbitrary units.}
\label{fig-6}
\end{figure}
\begin{figure}
\includegraphics[width=7cm]{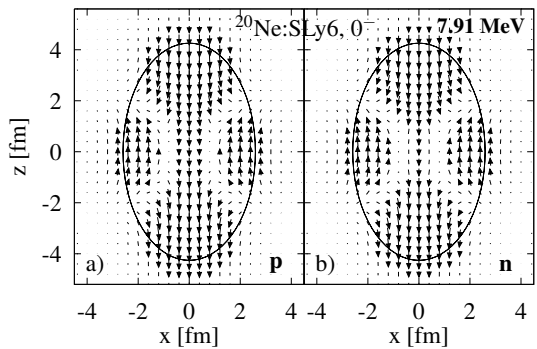}
\caption{The same as in Fig. 6 but
for K=0 7.91-MeV state.}
\label{fig-7}
\end{figure}

Like in $^{24}$Mg, TS and CS in $^{20}$Ne also exhibit the energy downshift with $\beta$.
However in $^{20}$Ne,
only K=0 CS becomes
the lowest excitation while  K=1 TS is preceded by other
dipole states ($\sim$7 and 8 MeV  at $\beta$=0.56 and
$\sim$6 and 7 MeV at $\beta$=0.72).
Moreover, unlike the $^{24}$Mg case, the CS lies below the TS.

\begin{table} 
\caption{The same as in Table I but for $^{20}$Ne at
$\beta$=0.56. The energy E is in [MeV],
the transition rate B(E3K) is W.u..}
\label{tab-2}       
\begin{tabular}{llllll}
\hline
E     & K & B(E3K)& main 2qp comp & F-pos & $\mathcal{N}$ \\\hline
7.91  & 0 & 45.6 & pp[220]$\uparrow$-[330]$\uparrow$ & F, F+5 & 0.21 \\
      &   &      & nn[220]$\uparrow$-[330]$\uparrow$ & F, F+5 & 0.16 \\
10.11 & 1 &    0.5&  pp[220]$\uparrow$+[330]$\uparrow$ & F, F+5 & 0.53 \\
      &   &       &nn[220]$\uparrow$+[330]$\uparrow$ & F, F+5 & 0.32 \\
\hline
\end{tabular}
\end{table}
\begin{figure} 
\includegraphics[width=7cm,clip]{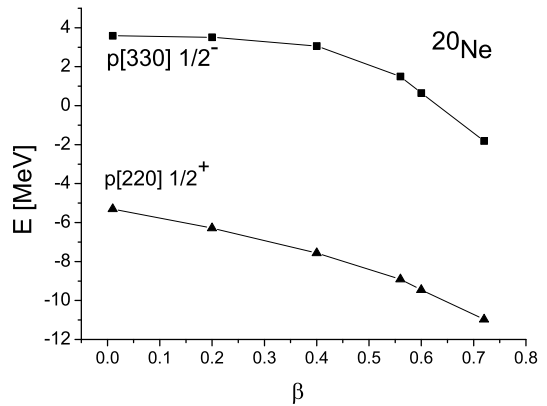}
\caption{The same as in Fig. 3 but for $^{20}$Ne.}
\label{fig-8}       
\end{figure}

These results can be partly understood by inspecting the structure
of the relevant states and their deformation dependence. Table II
shows that, in the toroidal K=1 state at 10.11 MeV and compressional
K=0 state at 7.91 MeV,  the major proton and neutron
2qp components embrace the same single-particle states, [220]$\uparrow$ and [330]$\uparrow$.
However, depending on their coupling into K=1 or K=0 configuration,
they produce the vortical toroidal or the irrotational compressional mode.
Like in $^{24}$Mg,  the major 2qp components in TS
exhaust most of the state norm (85\%). This once more confirms the previous findings
\cite{RW87,Dre15} that the toroidal mode is basically of mean field origin.
Instead, the compressional K=0 state at 7.91 MeV is much more collective.

In our QRPA calculations,  the forward amplitudes of conjugate
proton and neutron 2qp components given  in Tables I and II have
the same sign. This complies with the isoscalar character of the low-energy TS and
CS. Just in T=0 case, protons and neutrons contribute constructively into toroidal or
compression flows.

In figure 8, we show deformation dependence of the energies  of the
single-particle states [220]$\uparrow$ and [330]$\uparrow$
forming 2qp components in Table II. Both energies decrease
with $\beta$. At $\beta >$0.56,
the energy difference for these states somewhat decreases, which partly explains
the energy downshift  of the TS and CS. Note that the case of the collective CS is
more complicated and can hardly be treated by mere quasiparticle arguments.

It is interesting that the irrotational current for K=0 CS at 7.91
MeV, given in figure 7, is similar to the octupole flow for
the first $3^-$ state in $^{208}$Pb \cite{RW87}.  Moreover, in our calculations,
the $3^-$0 rotational state built on the collective
7.91-MeV band head results in a strong E30 transition from the ground
state (gs) band yielding $B(E30,0^+_{\rm{gs}} \to 3^-0)$=45.6
W.u. (see Table II). The latter can be explained by collectivity of the state
and by the fact that its  major 2q components fully obey
the selection rules for E30 transitions: $\Delta K=0$;
$\Delta N = \pm 1,\pm 3$;  $\Delta n_z = \pm1 ,\pm 3$;
$\Delta \Lambda = 0$ \cite{Ni65})  where
$[N n_z \Lambda]$K are Nilsson asymptotic quantum numbers.
The octupole features of the 7.91-MeV state
are not surprising since in deformed
nuclei the dipole and octupole excitations are coupled. One
may suggest that the compressional 7.91 MeV state is actually a
familiar octupole irrotational K=0 state. A similar conclusion
can be done for the compressional K=0 9.56-MeV state in $^{24}$Mg.

As seen from Tables I and II,  TS in $^{24}$Mg and $^{20}$Ne give very
different B(E31)-values: 11.7 W.u. and 0.5 W.u., respectively.
For the first glance, this is surprising since both states are non-collective
and their major 2qp components exhaust nearly the same amount of the state norm.
However, the difference in B(E3) can be easily understood using the selection rules
for E31 transitions: $\Delta K=\pm 1$; $\Delta N = \pm 1,\pm 3$;  $\Delta n_z = 0,\pm 2$;
 $\Delta \Lambda = 1$ \cite{Ni65}. It is easy to see that in $^{24}$Mg
the major TS components fully obey the selection rules and so this state exhibits
the large B(E31). Instead, in $^{20}$Ne, the major TS components
violate the rules for $n_z$ and $\Lambda$ and so the E31 transition is strongly
suppressed.

Following Tables I and II, the conjugate proton and neutron 2qp components
give similar but not precisely the same contributions to the state norm.
This means that predominately isoscalar TS and CS have a minor isovector admixture.
While the collective K=0 CS demonstrate the octupole features, the K=1
TS perhaps correspond to low-energy  isoscalar dipole states (LE-IDS) found in
light N=Z spherical doubly magic nuclei ($^{16}$O, $^{16}$Ca).  LE-IDS are characterized
by weak isospin-forbidden E1 transitions. They were widely explored some
time ago, see early studies  \cite{Mi75,Ca90} and recent detailed analysis  \cite{Papa11}.
Following \cite{Papa11}, these states can exhibit the toroidal flow. In our calculations,
K=1 TS in $^{24}$Mg and $^{20}$Ne also exhibit weak dipole transitions with
$B(E1, 0^+0_{\rm g.s.} \to 1^-1)$ = 2.52 $10^{-4}$ and 0.93 $10^{-4}$ $\rm{e^2 fm^2}$,
respectively. Perhaps these TS represent the realization of LE-IDS in strongly
deformed axial nuclei.

\section{Conclusions}

We analyzed the influence of strong axial quadrupole deformation on
the properties of individual low-energy E1(T=0) toroidal and
compressional states (TS and CS) in the N=Z nuclei $^{24}$Mg
($\beta_{\rm{exp}}$=0.605) and $^{20}$Ne ($\beta_{\rm{exp}}$=0.72).
The study was performed within the self-consistent quasiparticle
random-phase approximation approach using the Skyrme parametrization
SLy6. As shown in \cite{Ne_PRL18}, the large quadrupole deformation in
$^{24}$Mg causes a considerable energy downshift of TS and CS. As
a result, they become the energetically lowest dipole K=1 and K=0
excitations. Here we explain this effect by deformation dependence of
the energies of the major two-quasiparticle (2qp) components in TS and
CS.

Since the downshift effect is determined by deformation properties of
the particular 2qp configurations, it is not universal and in principle
can differ in other light deformed nuclei. Indeed, in $^{20}$Ne, the deformation
also yields the energy downshift of TS and CS. However, the TS is not
anymore the lowest dipole state even in K=1 channel. Instead CS is the
lowest dipole state lying below TS. Perhaps light deformed nuclei
with the lowest K=1 TS are rare. To our knowledge, they are limited by $^{24}$Mg
\cite{Ne_PRL18} and $^{10}$Be \cite{Kanada17}.

The calculation for $^{20}$Ne confirm the finding \cite{Ne_PRL18,Rep17,KvYb} that,
in axially deformed nuclei, the  E1 vortical toroidal strength dominates in K=1
low-energy spectra while the irrotational compressional strength prevails in K=0
channel. CS in $^{20}$Ne and  $^{24}$Mg are collective and remind  familiar
low-energy collective E30 octupole modes pertinent to deformed nuclei where
dipole and octupole excitations are mixed. Instead, TS are mainly formed by two
large conjugate proton and neutron 2qp components. So  the toroidal flow is basically
of mean-field origin. Perhaps TS in $^{20}$Ne and  $^{24}$Mg correspond
to low-energy T=0 dipole states with the isospin-forbidden E1 transitions, which were
earlier discussed for N=Z doubly-magic nuclei \cite{Mi75,Ca90,Papa11}. The TS could
be a realization of such states in strongly deformed nuclei. We plan to consider
this point in next studies.

\vspace{0.5cm}
V.O.N. thanks Profs. P. von Neumann-Cosel, J. Wambach and V.Yu. Ponomarev
for valuable discussions. The work was partly supported by the
Heisenberg - Landau (Germany - BLTP JINR), and Votruba - Blokhintsev (Czech Republic
- BLTP JINR) grants. A.R. is grateful for support from Slovak Research and Development
Agency (Contract No. APVV-15-0225) and Slovak grant agency VEGA (Contract No. 2/0129/17).
J.K. appreciates the support of the research plan MSM 0021620859 (Ministry of Education of the
Czech Republic) and the Czech Science Foundation project P203-13-07117S.%

\end{document}